\documentstyle[12pt,aaspp4,overcite,flushrt,here]{article}
\tighten

\renewcommand{\thebibliography}[1]{\clearpage\subsection*{REFERENCES}\list
 {\arabic{enumi}.}{\settowidth\labelwidth{[#1]}\leftmargin\labelwidth
 \advance\leftmargin\labelsep
 \usecounter{enumi}}
 \def\newblock{\hskip .11em plus .33em minus .07em}
 \sloppy\clubpenalty4000\widowpenalty4000
 \sfcode`\.=1000\relax}

\def\aap{{\sl Astr.~Astrophys.\/}}
\def\aj{{\sl Astr.~J.\/}}
\def\apj{{\sl Astrophys.~J.\/}}
\def\apjlett{{\sl Astrophys.~J.\/}}
\def\apjsupp{{\sl Astrophys.~J.\ Suppl.\/}}
\def\baas{{\sl Bull.~Amer.~Astr.~Soc.\/}}
\def\mnras{{\sl Mon. Not. R. astr. Soc.\/}}
\def\nat{{\sl Nature\/}}
\def\pasp{{\sl Publ.~astr.~Soc.~Pacif.\/}}
\def\ang{\thinspace{\rm \AA}}
\def\approxlt{\lower.2em\hbox{$\buildrel < \over \sim$}}
\def\civ{{\rm C}\thinspace{\sc{iv}}}
\def\erg{${\rm erg\ cm}^{-2}\ {\rm s}^{-1}$}

\def\ergfnu{{\rm erg\ cm}^{-2}\ {\rm s}^{-1}\ {\rm Hz}^{-1}}
\def\ha{\ifmmode {{\rm H}\alpha}
        \else {H$\alpha$}\fi}
\def\hnought{\ifmmode H_0
    \else $H_0$\fi}

\def\kms{~{\rm km\ s}^{-1}}
\def\la{\ifmmode {{\rm Ly}\alpha}
        \else {Ly$\alpha$}\fi}
\def\msun{{$M_{\odot}$}}
\def\oii{[O\thinspace{\sc{ii}}]}
\def\p.{^{\prime\prime}\kern-2.1mm .\kern+.6mm}
\def\pone{^{\prime}\kern-1.05mm .\kern+.3mm}
\def\qnought{\ifmmode q_0
    \else $q_0$\fi}
\def\sqr#1#2{{\vcenter{\hrule height .#2pt
        \hbox{\vrule width .#2pt height#1pt \kern#1pt
                \vrule width .#2pt}
        \hrule height.#2pt}}}
\def\square{\mathchoice\sqr56\sqr56\sqr{4.1}5\sqr{3.5}5}
\def\ten#1{\ifmmode 10^{#1}
    \else $10^{#1}$\fi}

\begin{document}

\title{Detection of Lyman-{$\boldmath \alpha$} emitting galaxies at 
  redshift {$\boldmath z$} = 4.55}
\author{Esther M. Hu\altaffilmark{\ast,\dagger}} 
\and
\author{Richard G. McMahon\altaffilmark{\ddagger}}
\altaffiltext{1}{Institute for Astronomy, University of Hawaii, 2680 Woodlawn 
  Drive, Honolulu, Hawaii 96822, USA}
\altaffiltext{2}{Visiting Astronomer, W. M. Keck Observatory, 
  California Association for Research in Astronomy.}
\altaffiltext{3}{Institute of Astronomy, University of Cambridge, Madingley
  Road, Cambridge CB3\thinspace{0HA}, UK}

\vskip1.9in
\centerline{To appear in Nature}
\newpage

%
{\bf Studies of the formation and early history of galaxies have been
hampered by the difficulties inherent in detecting faint galaxy populations
at high redshift.  As a consequence, observations at the highest redshifts
(3.5 $\boldmath < z < $ 5) have been restricted to objects that are
intrinsically bright. These include quasars, radio galaxies, and some
Ly\mbox{\boldmath$\alpha$}-emitting objects\cite{br1202,fontana96,petit96}
that are very close to (within \mbox{\boldmath$\sim$ 10} kpc)
--- and appear to be physically associated with --- quasars.  But the
extremely energetic processes which make these objects easy to detect also
make them unrepresentative of normal (field) galaxies.  Here we report the
discovery of two Ly{\mbox{\boldmath$\alpha$}}-emitting galaxies at redshift
\mbox{\boldmath{$z =$}} 4.55, which are sufficiently far from the nearest
quasar (\mbox{\boldmath$\sim$ 700} kpc) that radiation from the quasar is
unlikely to provide the excitation source of the Ly{\mbox{\boldmath$\alpha$}}
emission.  Instead, these galaxies appear to be undergoing their first
burst of star formation, at a time when the Universe was less than one
billion years old.}

Searches for high-$z$ field galaxies have progressed from several
directions.  Extensive spectroscopic surveys of field
galaxies\cite{songaila_3,lilly_1,glazebrook_b,large_sample} now show
numerous random field galaxies at $z>1$, with evidence that many of these
are in the process of intense star formation\cite{gal_form}.  Searches for
the galaxies associated with quasar absorption lines are also beginning to
turn up normal galaxies in this redshift range\cite{dic95}.  At $z>1.7$,
where the \oii\ emission line at 3,727\ang\ is no longer detectable in the
optical window, we are beginning to find objects on the basis of their UV
absorption lines\cite{hzsb,haw167,yee96,stei96}.  These objects may be
starbursting galaxies, in some cases with active galaxy (AGN) activity.

At redshifts near 3, absorption-line objects should also show distinct
continuum color breaks as the limit for the hydrogen Lyman series (912
\AA\ in the rest frame) becomes observable at ultraviolet wavelenths.  The
objects identified by color breaks and UV absorption features to lie at
$3.0 < z < 3.5$ [ref.~\citen{stei96}] should be relatively evolved
galaxies, whose bright UV continuua are produced by massive stars which
chemically enrich these systems, giving rise to metal-line absorption
features and dust, making \la\ emission weak or undetectable.

Galaxies at earlier stages of this process, prior to the formation of dust,
may have much stronger \la\ emission relative to the stellar continuum.
Such objects, which are faint in the continuum, may be hard to pick out
with color break techniques but relatively easy to find by \la\ searches.
A small number of such objects have been seen in the fields of
$z\sim2\to3.5$ quasars, which are well separated from the quasar and which
may represent neighboring galaxies either with AGN\cite{low91} or with
intense star formation that can excite the \la\ emission line but without
so much dust that the line is suppressed\cite{gia94,macch93,francis}.  The
latter class of objects, which are relatively unobscured by foreground
absorption, may provide the most direct estimates of star formation at the
earliest epochs.  It appears that the quasar may mark the sites of such
formation and thus identify a redshift for targeted wavelength
searches\cite{djorg85}.  The present paper reports detection of such
objects at by far the highest redshift yet in the field of the $z=4.55$
quasar BR2237--0607\cite{sto96}.

In order to search for high-$z$ objects in the fields of the $z>4$
quasars, we first obtained exposures through a narrow-band filter centered
on the quasar's redshifted \la\ emission, and through a nearby broader
continuum, with a Tektronix 2048$^2$ camera on the Univ.\ of Hawaii's
2.2-m telescope. Five $z>4$ quasar fields have now been imaged, but the
spectroscopic follow-up is complete only for BR2237--0607, which we
describe here.  Narrow-band and continuum images on this field are shown
in Fig.~1.  All objects with a significant excess in the \la\ bandpass
were considered candidate \la\ emitters, and this initial search turned up
10 objects in the 19.25$\square'$ field surrounding BR2237--0607, a
subsample of which is circled in Fig.~1. However, many of these candidate
objects correspond to lower redshift galaxies whose emission lines
coincidentally match the redshifted \la\ wavelength, or to objects whose
continuum shape makes them significantly brighter in the narrow-band
filter.  Therefore, in order to proceed further we obtained simultaneous
spectra of the candidate objects using the multi-object LRIS
spectrograph\cite{lris} on the Keck 10-m.\ telescope.

Two-dimensional spectra for the three candidate objects of Fig.~1 are shown
in Fig.~2, with the wavelength region of the narrow-band and reference
continuum band filters marked at the top of the plot.  As expected, most of
the candidate objects were identified as $z < 1$ galaxies.  We show in this
figure, for instance, the spectrum of the low-redshift emission-line
galaxy, LA10, where the strong emission is identified with the
\oii\ 3,727\ang\ line instead of \la, for a redshift of 0.802.  The spectra
show various features that are hallmarks of this class of interloper,
namely: much stronger continuum, including clearly detected flux from the
region below 5,000\ang\ which would correspond to the Lyman limit break at
the rest-frame 912\ang\ photoionization threshold of neutral hydrogen for
$z\sim4.6$ objects; and the offset wavelength position near the edge of the
narrow passband filter.  However, the two objects denoted here as LA1 and
LA2 exhibit only an extremely strong single emission line close to the
central wavelength of the narrow-band filter together with an extremely
weak continuum.  The insets for Fig.\ 1 show that for LA1 the continuum
is undetected even in an extremely deep 1-hour $I$-band image obtained
with LRIS on Keck, while a weak diffuse structure is seen in this band
for LA2, whose continuum is also faintly detected in the spectroscopic
data.  The Kron-Cousins $I\/$ magnitude of LA2 is 24.69 and it is
detected at the 4$\sigma$ level in the Keck $I$-band image (Table 1).
Spectra of these two objects are shown in Fig.~3.  A useful measure of
the relative strength of line to continuum is the equivalent width, which
is defined as the width of a rectangle with the same area as an emission 
(or absorption) feature, with a height equal to the continuum level above 
zero.  The observed equivalent widths ($>700$ \AA\ [LA2] and $>1,300$ \AA\ 
[LA1]) are quoted only as lower limits because of the weakness of the 
continuum, but are too large for these lines to be \oii\ or
\ha\cite{kenn83,kenn92,scl} --- an interpretation which is also ruled out
by the absence of any other strong emission lines.  This leaves little
alternative but to identify these objects as \la\ emitters at the redshift
of the quasar.  Their coordinates and properties may be found in Table~1.
The equivalent width of object LA10 (123 \AA) is consistent with the
expected range of values for \oii\ emission\cite{songaila_3,large_sample}.

The separations from the quasar of $117''$ for LA1 and $105''$ for LA2 are
too large for the quasar to have a significant role in exciting the
objects\cite{hu91} (1$'' = 7.1\ h^{-1}$ kpc at $z=4.55$, where $h =
\hnought/100\kms$ Mpc$^{-1}$ and \qnought=0.5) and it appears that the
objects must be either internally excited or ionized by the general
metagalactic UV flux\cite{scl}.  As can be seen from the inset to Fig.~1,
LA1 is quite compact and could have an AGN component, although there is no
sign of emission from \civ\ at 1,549\ang.  However, LA2 is quite diffuse in
both $I$ and \la, and most likely is excited by star formation.  An
alternative possibility is that the \la\ emission is produced by surface
ionization of the gas cloud by the metagalactic UV flux.  The metagalactic
ionizing flux at these wavelengths is currently estimated\cite{williger94}
as $1-3 \times\ten{-22}\ \ergfnu$ sr$^{-1}$, which would produce an observed
\la\ flux of $\sim2 \times$ \ten{-17}\ \erg\ from an object with a 2$''$
angular size at $z=4.55$ (ref.~\citen{scl}) if there is one \la\ photon
emitted per ionization and we see only the forward side of the cloud, so it
is possible we are seeing a component due to ionized gas clouds.  However,
this would provide no explanation for the continuum light in LA2.

The rest frame equivalent widths of the systems are around $>130$\ang\ (LA2)
and $>240$\ang\ (LA1), which are marginally consistent with stellar
excitation for an initial mass function dominated by massive
stars\cite{charl93}.  Following the discussion of the previous paragraph,
the \la\ may arise from a combination of internal and external ionization.
If the observed emission were primarily due to stars, and there were no
internal scattering and extinction, then the luminosity of
$3\times\ten{42}$ $h^{-2}$ \erg\ (\qnought=0.5) would correspond to a star
formation rate in solar masses (\msun) per year of $\sim3\ h^{-2}$
\msun\ yr$^{-1}$, where we use Kennicutt's (ref.~\citen{kenn83}) relation
between \ha\ luminosity and star formation rate (SFR) of SFR = $L($\ha)
$\times\ 8.9\ \times\ten{-42}$ erg s$^{-1}$ \msun\ yr$^{-1}$, and assume a
ratio of \la\ to \ha\ (8.7) that applies for Case B
recombination\cite{brock}.  Given the Hubble time at this redshift of
$7\times\ten{8}\ h^{-1}$ yr (\qnought=0.5) the integrated amount of star
formation is small compared to that of a `normal' galaxy with
$6\times\ten{10}\ h^{-1}$ \msun\ of stars (a so-called $L^*$ galaxy).

Because of the targeted nature of the search it is hard to estimate from
the present data whether such objects may be common in the general field or
whether they are preferentially found around quasars.  Observations, 
currently in progress, of additional quasars and blank field regions
should answer this question.


\vskip0.1in
\centerline{\bf Acknowledgements}

We thank T.\ Bida, R.\ Campbell, T.\ Chelminiak, and B.\ Schaefer
for their assistance in obtaining the observations, which would not have
been possible without the LRIS spectrograph of J.\ Cohen and B.\ Oke. 
Research at the University of Hawaii was supported by the State of
Hawaii and by NASA.  E.M.H.\ would also like to gratefully acknowledge
a University Research Council Seed Money grant.  R.G.M. acknowledges
the support of the Royal Society.

\newpage

\begin{figure}[H]
\caption{Central $3\pone5  \times 3\pone5$ region showing the $z=4.55$
quasar BR2237--0607 and emitting objects LA1, LA2, and LA10 in a
narrow-band filter (central wavelength 6,741 \AA, 51\AA\ bandpass) centered
on the quasar's redshifted \la\ emission [left panel] and through a nearby
broad, line-free continuum filter (7,500 \AA, 698 \AA\ bandpass) [right
panel].  Inset panels show the region around identified \la\ emitters LA1
and LA2, both in line and continuum, and enlarged by a factor of two. Here
the continuum is a deep 1 hr $I$\ band taken at Keck.  LA2 has a weak
continuum (Fig.~2), but is clearly extended in both the \la\ and $I$-band
images.  The continuum for LA1 is not detected in the 1-hr $I$-band
exposure at a $1\sigma$ magnitude of 26.15.  The data were taken as a
series of sky noise-limited integrations, with an offset step of $10''$
between successive frames, and a median sky exposure was generated from the
on-field exposures for each night and used to correct for non-uniformities
in detector response (flat-field).  Spectrophotometric standards (Feige 15,
Feige 110, Kopff 27, BD+28 4211)\cite{massey88,stone} for the narrow-band
data and a combination of spectrophotometric (Feige 110, BD+28
4211)\cite{stone} and Landolt standards (in the fields of SA 92-248,
PG1633+099, PG0231+051)\cite{landolt92} for the continuum data were used to
calibrate the data, and were taken in observations both before and after
the target exposures.  The composite image quality was $1\p.0$ in both the
20.8 hr narrow-band image and in the 1 hr 7,500 \AA\ continuum image, and
$0\p.8$ in the deep 1 hr Keck $I$-band image.}
\end{figure}
\begin{figure}[H]
\caption{LRIS spectra of the 3 candidates from Fig.~1 in the BR2237--0607
field.  Observations were made with a multi-slit mask over the quasar field,
with a slit width of $1\p.4$, spectral resolution of $\sim17$\ang, and
spatial sampling of $0\p.215$/pixel.  The results of 3 hrs exposure in a
single mask setting are shown as two-dimensional spectra, with night sky
lines subtracted from the spectrum, and the displaced starting wavelengths
reflect the offsets in slit position for different objects in the field.  The
spatial extent of each spectrum is $\sim6\p.5$ --- roughly the diameter
encircling the objects in Fig.~1.  At the top of the plot we indicate the
wavelength regions covered by the narrow-band filter (central wavelength 6,741
\AA, 51 \AA\ bandpass), which is centered near the wavelength of the quasar's
\la\ emission, and the reference continuum filter (central wavelength
7,500 \AA, 698 \AA\ bandpass).  Objects LA1 and LA2, which are well centered
in the narrow-band filter, have observed equivalent widths in excess of 700
\AA\ in the emission line, and thus must almost certainly be due to
redshifted \la.  No other features are detected, apart from a faint continuum
in LA2.  Object LA10 is an \oii\ emitter at $z=0.802$ with a line equivalent
width of $\sim 123$ \AA, and can be easily identified, even in the absence of
other emission lines, by its strong continuum, extending blueward of
5,000 \AA\ (roughly, the wavelength of the break in the Lyman limit absorption
for $z\sim4.6$ objects).  The asymmetric placement of the emission line is
another clue of a low-$z$ interloper.}
\end{figure}
\begin{figure}[H]
\caption{Extracted spectra for LA1 and LA2 plotted in the rest wavelength
frame.  The spectra represent a total of 4.7 hrs integration with the LRIS
spectrograph on Keck.  The emission wavelengths agree to within
$\sim$2.4\ang\ ($\sim100\kms$ at $z=4.55$; $\sim0.4$ \AA\ in the
rest frame) for these two objects. LA1 is
not resolved at 3.1 \AA\ in the rest frame, and the width of LA2 is most 
likely due to the extended emission structure across the slit.}
\end{figure}
\end{document}